\documentclass[12pt, preprint]{aastex}

\usepackage{graphicx}
\usepackage{amssymb}
\usepackage{epstopdf}
\usepackage{layout}
\usepackage{url}
\usepackage{subfigure}
\slugcomment{Accepted for Publication in {\it The Astrophysical Journal}}

\shorttitle{X-Ray and Gamma-Ray Polarization in Blazars}
\shortauthors{H. Zhang \& M. B\"ottcher}

\begin{document}

\title{X-Ray and Gamma-Ray Polarization in Leptonic and Hadronic Jet Models of Blazars}

\author{H. Zhang\altaffilmark{1} and M. B\"ottcher\altaffilmark{2,1}}

\altaffiltext{1}{Astrophysical Institute, Department of Physics and Astronomy, \\
Ohio University, Athens, OH 45701, USA}

\altaffiltext{2}{Centre for Space Research, North-West University, Potchefstroom,
2520, South Africa}

\begin{abstract}
We present a theoretical analysis of the expected X-ray and $\gamma$-ray 
polarization signatures resulting from synchrotron self-Compton emission
in leptonic models, compared to the polarization signatures from proton synchrotron
and cascade synchrotron emission in hadronic models for blazars. Source parameters
resulting from detailed spectral-energy-distribution modeling are used to calculate
photon-energy-dependent upper limits on the degree of polarization, assuming a 
perfectly organized, mono-directional magnetic field. In low-synchrotron-peaked
blazars, hadronic models exhibit substantially higher maximum degrees of X-ray and 
gamma-ray polarization than leptonic models, which may be within reach for existing
X-ray and $\gamma$-ray polarimeters. In high-synchrotron-peaked blazars (with 
electron-synchrotron-dominated X-ray emission), leptonic and hadronic models
predict the same degree of X-ray polarization, but substantially higher maximum
$\gamma$-ray polarization in hadronic models than leptonic ones. These predictions are 
particularly relevant in view of the new generation of balloon-borne X-ray polarimeters 
(and possibly GEMS, if revived), and the ability of {\it Fermi}-LAT to measure 
$\gamma$-ray polarization at $< 200$~MeV. We suggest observational strategies combining
optical, X-ray, $\gamma$-ray polarimetry to determine the degree of ordering of the
magnetic field and to distinguish between leptonic and hadronic high-energy emission. 
\end{abstract}


\section{Introduction}

While measurements of synchrotron polarization of the radio and optical emission 
from relativistic jet sources (blazars, radio galaxies, gamma-ray bursts) have 
become a standard way of assessing the degree of order and direction of magnetic 
fields, the polarization of high-energy (X-ray and $\gamma$-ray) emission has so 
far remained largely unexplored. However, there are several projects underway to 
develop balloon- and satellite-borne X-ray polarimeters: Current balloon-borne
X-ray polarimetry experiments include PoGoLite \citep[2 -- 100~keV;][]{Pearce12}, 
X-Calibur \citep[20 -- 80~keV;][]{Beilicke12}, and POLAR \citep[50 -- 500~keV;][]{Orsi11}.
As satellite-borne instruments, SPI and IBIS on board the INTEGRAL satellite have 
already been used successfully to constrain the hard X-ray / soft $\gamma$-ray 
polarization from gamma-ray bursts \citep{Dean08,Forot08}, and design studies for the 
upcoming ASTRO-H mission suggest that it may also be able to detect polarization 
in the 50 -- 200~keV energy band \citep{Tajima10}. Estimates of the minimum
detectable degree of polarization for most of these instruments range around 10~\%
for moderately bright X-ray sources. Unfortunately, the Gravity and 
Extreme Magnetism SMEX \citep[GEMS; 2 -- 10~keV;][]{Swank10} was recently cancelled 
by NASA, in spite of being rather far along in its developing phase. It has also
been suggested that the Large Area Telescope (LAT) on-board the {\it Fermi}
Gamma-Ray Space Telescope may be able to detect $\gamma$-ray polarization in the
energy range $\sim 30$ -- 200~MeV when considering pair-conversion events occurring 
in the Silicon layers of the detector, by taking advantage of the polarization-dependent
direction of motion of the electron-positron pairs produced in the $\gamma$-ray --
pair conversion process \citep{Buehler10}. For bright $\gamma$-ray sources, degrees 
of polarization down to $\sim 10$~\% may be detectable.

On the theory side, the general formalism for calculating X-ray and $\gamma$-ray
polarization has been well developed. It is well-known that linear polarization
arises from synchrotron radiation of relativistic charged particles in ordered
magnetic fields, while Compton scattering off relativistic electrons will reduce
the degree of polarization of the target photon field, without entirely destroying
it. Compton scattering of unpolarized target photon fields by isotropic distributions
of electrons (and positrons) will always result in un-polarized Compton emission.
The well-known formalism for calculating synchrotron polarization can be found, e.g., 
in the text book by \cite{RL85}. A formalism for evaluating the polarization of 
Compton scattered radiation in the Thomson regime was developed by \cite{BCS70}
and applied specifically to synchrotron self-Compton (SSC) emission by \cite{BS73}.
More recently, \cite{Krawczynski12} provided a general Monte-Carlo based framework
for the evaluation of polarization signatures in relativistic environments in both
Thomson and Klein-Nishina regimes, verifying that the expressions of \cite{BCS70}
and \cite{BS73} are valid in the Thomson regime. 

However, while the general framework for the evaluation of polarization signatures
from different radiation mechanisms exists, it has so far not been applied to 
realistic representations of the high-energy emission from relativistic jets of 
active galactic nuclei (AGN), which are the  most numerous source class in the 
{\it Fermi}-LAT energy range. The $\gamma$-ray brightest AGN detected by {\it Fermi}-LAT 
are blazars, i.e., radio-loud (jet-dominated) AGN in which the jet points at a small 
angle with respect to our line of sight.
Example results for the synchrotron and SSC polarization in blazar jets
have been presented by \cite{Poutanen94}, indicating that substantial ($> 30$~\%) 
polarization may result from these radiation mechanisms in the case of perfectly
ordered magnetic fields. However, his calculations were restricted to pure power-law 
electron spectra and only a very specific choice of parameters, with no direct 
connection to the observed spectral energy distributions (SEDs) of blazars.
The SEDs of blazars are dominated by non-thermal emission across the entire electromagnetic
spectrum, with two broad components. Depending on the peak frequency of the
low-frequency component (generally agreed to be synchrotron radiation from
relativistic electrons), they are sub-divided between Low-Synchrotron-Peaked
(LSP) blazars (consisting of flat-spectrum radio quasars [FSRQs] and low-frequency
peaked BL~Lac objects [LBLs]), Intermediate-Synchrotron-Peaked (ISP) blazars
(generally intermediate BL~Lac objects [IBLs]), and High-Synchrotron-Peaked
(HSP) blazars (which are exclusively high-frequency-peaked BL~Lac objects 
[HBLs]). 

The mechanism producing the high-energy (X-ray through $\gamma$-ray) emission
in blazars is still under debate. Both leptonic and hadronic models are currently
still viable and are generally able to produce acceptable fits to the SEDs of
most blazars \citep[for a review of blazar emission models see, e.g.,][]{Boettcher10}.
Additional information, such as variability or polarization, is therefore needed
to distinguish between leptonic and hadronic emission scenarios. 

In this paper, we evaluate the expected X-ray and $\gamma$-ray polarization
signatures in both leptonic and hadronic emission models for blazars, considering
all sub-classes (LSP, ISP, and HSP) of blazars. In \S \ref{theory}, we briefly 
describe the formalism used to evaluate synchrotron and SSC polarization in this 
work, and the leptonic and hadronic blazar emission models considered. In \S 
\ref{results}, we present our results comparing the frequency-dependent polarization
signatures of leptonic and hadronic blazar models for blazars of all
sub-classes. These results are used in \S \ref{diagnostics} to develop 
observational diagnostics based on optical, X-ray and $\gamma$-ray polarimetry
to confidently distinguish between leptonic and hadronic emission models.
We summarize in \S \ref{summary}.

\section{\label{theory}Calculation of Synchrotron and Synchrotron-Self-Compton 
Polarization}

In this section, we will describe the formalism used in this paper to evaluate 
the degree of polarization from synchrotron and Compton scattering, and the
specific leptonic and hadronic blazar models used to provide input parameters
for the polarization calculations. All results shown here are based on the
assumption of a perfectly ordered, mono-directional magnetic field, oriented
perpendicular to the line of sight. This configuration produces the maximum 
possible degree of polarization both for synchrotron and SSC emission. 
Thus, our results represent upper limits to the actually expected 
polarization from emission regions with partially dis-ordered magnetic fields. 
As we will elaborate in Section \ref{diagnostics}, the observed degree of 
polarization in frequency ranges in the SED that can be confidently ascribed 
to synchrotron radiation can then be used to re-normalize our results to find 
the actually expected degree of X-ray and $\gamma$-ray polarization.

All particle distributions, on which we base our calculations, are
assumed to be isotropic in the co-moving frame of the emission region, as
is routinely done in blazar emission models. This is justified by very
efficient pitch-angle scattering on small-scale magnetic turbulence (with
small amplitudes $\delta B / B_0 \ll 1$, where $B_0$ is the ordered
background magnetic field) which is expected to be present in the
highly relativistic blazar environment. For a more detailed discussion
of this aspect, see, e.g., \cite{bms97}.

In general terms, we will evaluate the radiation powers $P_{\parallel}$ 
and $P_{\perp}$ of radiation with electric-field vectors parallel and 
perpendicular, respectively, to the projection of the magnetic field onto 
the plane of the sky (which is identical to the actual magnetic field in 
the configuration chosen for our calculations). The degree of polarization 
$\Pi$ is then evaluated as

\begin{equation}
\Pi (\omega) = {P_{\perp} (\omega) - P_{\parallel} (\omega) \over 
P_{\perp} (\omega) + P_{\parallel} (\omega) }
\label{Pidef}
\end{equation}

\subsection{\label{SySSC}Synchrotron and Synchrotron-Self-Compton Polarization}

For the case of synchrotron emission, $P_{\parallel} (\omega)$ and $P_{\perp} 
(\omega)$ are evaluated by integrating the single-particle powers $P_{\parallel} 
(\omega, \gamma)$ and $P_{\perp} (\omega, \gamma)$ from Eqs. (6.32a,b) in \cite{RL85} 
over the electron spectrum $N_e(\gamma)$, e.g., $P_{\parallel, \perp} (\omega) = 
\int_1^{\infty} P_{\parallel, \perp} (\omega, \gamma) \, N_e(\gamma) d\gamma$. 
This procedure is valid for arbitrary particle spectra (not only the most
commonly considered power-law case) and is followed for synchrotron emission from 
all relevant species, i.e., electrons in both leptonic and hadronic models as well as
protons and electron-positron pairs from cascades in hadronic models. Our calculations
properly account for the mass difference in the case of proton synchrotron
\citep[see, e.g.][]{aharonian00}.
It is well known that a power-law spectrum of electrons $N_e (\gamma) \propto 
\gamma^{-p}$ (producing a power-law synchrotron radiation spectrum $F_{\nu} 
\propto \nu^{-\alpha}$ with $\alpha = (p - 1)/2$) will result in a polarization 
degree of \citep{RL85}

\begin{equation}
\Pi_{\rm powerlaw} = {p + 1 \over p + 7/3} = {\alpha + 1 \over \alpha + 5/4}
\label{Pipowerlaw}
\end{equation}

For Compton scattering, we follow the formalism of \cite{BS73}. 
We only need to calculate the polarization for SSC emission since X-ray and $\gamma$-ray
emission from Compton scatering of radiation fields external to the $\gamma$-ray emission
region (EC = External Compton) is expected to be unpolarized, and can simply be added as
a radiation component with $P_{\parallel}^{\rm EC} (\omega) = P_{\perp}^{\rm EC} 
(\omega)$. 

Since the expressions for SSC polarization given by \cite{BS73} are only valid in 
the Thomson regime, we must first confirm their validity in the considered energy
range of $\sim 0.1$~keV -- 500~MeV, for which we will show polarization results
in Section \ref{results} in order to encompass the energy ranges covered by X-ray 
polarimeters and {\it Fermi}-LAT's polarization capabilities. Here and throughout 
the paper, we define a dimensionless photon energy $\epsilon \equiv h \nu / (m_e c^2)$. 
Doppler boosting from the co-moving to the observer's frame is described by the
Doppler factor $\delta \sim 10$ for typical blazar sources. In the Thomson regime, 
the observed Compton-scattered photon energy $\epsilon_C^{\rm obs}$ is given by 
$\epsilon_C^{\rm obs} = \gamma ^2 \delta \epsilon_s$, where $\epsilon_s$ is 
dimensionless target photon energy in the co-moving frame of the $\gamma$-ray
emission region. 
For an observed photon energy of $E_C^{\rm obs} = 500$~MeV, $\epsilon_C^{\rm obs} 
\sim 1000$. In the 
SSC process, the target synchrotron photons have a typical photon energy around 
$\hbar \omega \sim eV$, so that $\epsilon_s \sim 10^{-6}$. Hence, $\gamma \epsilon_s 
= \sqrt{\epsilon_C^{\rm obs} \epsilon_s \delta^{-1}} \sim 0.01 \ll 1$. 
We can therefore safely work in the Thomson regime, for which
\cite{Krawczynski12} has verified with detailed Monte-Carlo simulations that the
expressions of \cite{BS73} provide an excellent description of the polarization
signatures.

\begin{figure}[ht]
\plotone{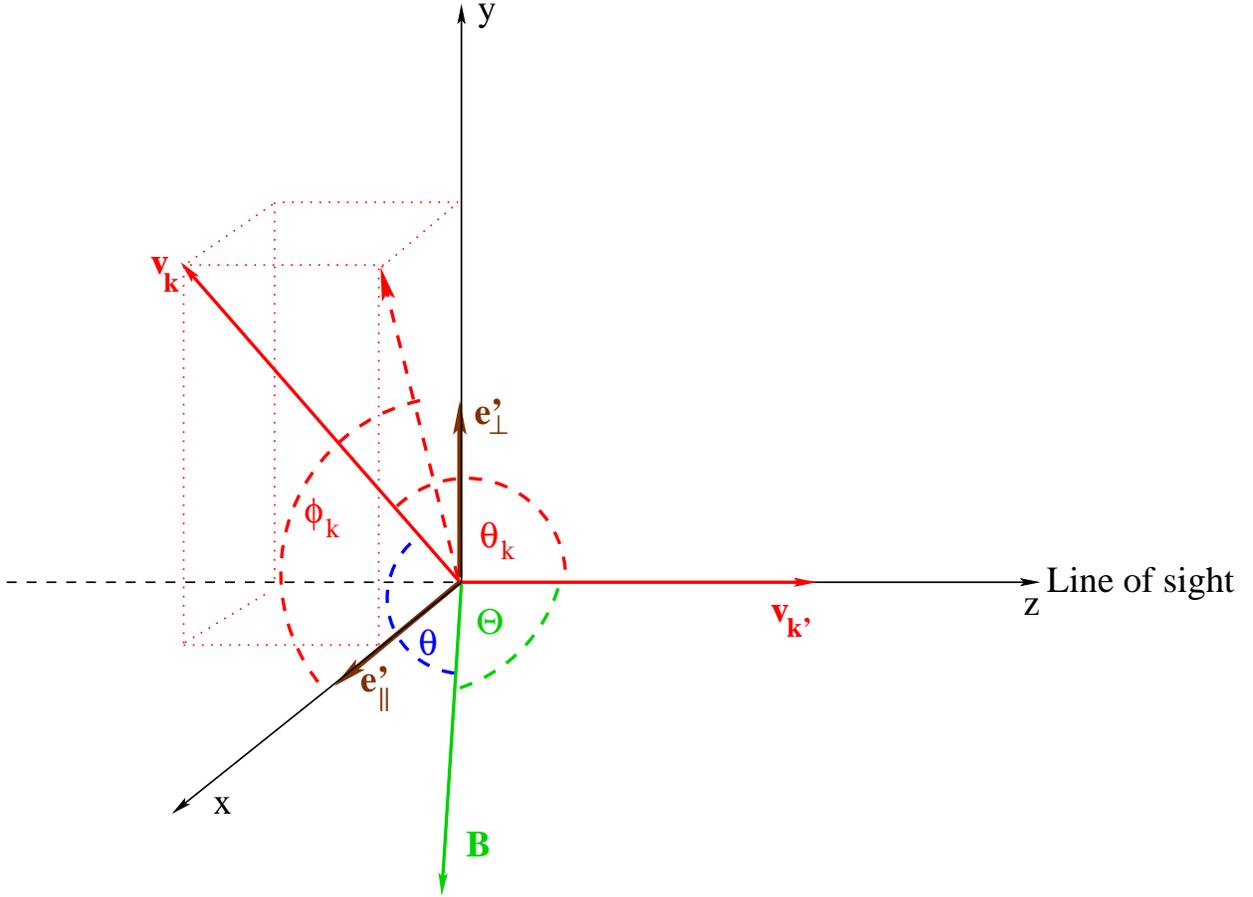}
\caption{Sketch of the scattering geometry to illustrate the definition of
angles. The co-ordinate system is chosen such that the magnetic field {\bf B} 
lies in the (x-z) plane. See the electronic edition of the Journal for a color 
version of this figure.\label{geometry}}
\end{figure}

Let $ \bf k$, $\epsilon$ and $ \bf k'$, $\epsilon'$ be the momentum and the frequency 
of the photons before and after scattering, with unit vectors ${\bf v_k} = 
{\bf k}/\vert{\bf k}\vert$ and ${\bf v_{k'}} = {\bf k'}/\vert{\bf k'}\vert$. 
In the Thomson regime, the power 
emitted at frequency $\epsilon'$ resulting from scattering of photons whose original 
polarization direction is $\bf e$ (taken as being perpendicular to the magnetic field
for synchrotron target photons), scattered into two orthogonal polarization 
directions ($\bf e'$ parallel or perpendicular to the projection of the magnetic
field onto the plane orthogonal to $\bf k'$), $P_{\parallel}^{\rm SSC}$ and 
$P_{\perp}^{\rm SSC}$, are given by \citep[][converted to c.g.s units]{BS73}
\begin{mathletters}
\begin{eqnarray}
P_{\parallel}^{\rm SSC} (\epsilon') & = & \pi (\frac{e^2}{4\pi})^2  \frac{c}{m_e c^2} \epsilon' 
\int  \frac{d\epsilon}{\epsilon} d\Omega_k \; E_{min} n(\epsilon) q(\vartheta) \cdot 
( Z_{\parallel} (\Sigma _1 + \Sigma _2 )+\Sigma _2) \\
P_{\perp}^{\rm SSC} (\epsilon') & = & \pi (\frac{e^2}{4\pi})^2  \frac{c}{m_e c^2} \epsilon' 
\int  \frac{d\epsilon}{\epsilon} d\Omega_k \; E_{min} n(\epsilon) q(\vartheta) \cdot 
( Z_{\perp} (\Sigma _1 + \Sigma _2 )+\Sigma _2)
\label{PSSC}
\end{eqnarray}
\end{mathletters}
where
\begin{equation}
E_{min} = \sqrt{\frac{\epsilon'}{2\epsilon (1-\cos \theta _k)}} 
\label{Emin}
\end{equation}
is the minimum electron energy required for scattering of a photon from $\epsilon$ 
to $\epsilon'$, 

\begin{equation}
Z_{{\bf e'}} = \left( {\bf e \cdot e'} + 
\frac{{\bf (v_k \cdot e' )(v_{k'} \cdot e)}}{1-\cos \theta _k} \right)^2
\label{Zdef}
\end{equation}

and we have defined the solid angle of the direction of the photon before
scattering, $\bf k$, as
\begin{equation}
d\Omega _{\bf k} = d \cos \theta _k d \varphi _k 
\label{Omegak}
\end{equation}

$\vartheta$ is defined as the angle between the magnetic field and $\bf k$, which
can be related to the angle $\Theta$ between the magnetic field and the line of 
sight ($\bf k'$) through
\begin{equation}
\cos \vartheta = \cos \Theta \cos \theta _k + \sin \Theta \sin \theta _k \cos \varphi _k 
\label{theta}
\end{equation}
See Figure \ref{geometry} for an illustration of the angle definitions. 

The synchrotron photon distribution $n({\bf k})$ has been separated into an energy
spectrum and an angle-dependent function,
\begin{equation}
n({\bf k}) = \frac {n(\epsilon)q(\vartheta)}{\epsilon^2}
\label{nepsilon}
\end{equation}
where the angle-dependence is chosen as $q(\vartheta) \propto \sin^{p + 1 \over 2}
\vartheta$ with $p$ being the local spectral index of the underlying electron spectrum,
and the synchrotron spectrum $n(\epsilon)$ is calculated self-consistently, using
the full expressions of \cite{RL85} for the given electron spectrum.

\begin{eqnarray}
\Sigma_1 & = & \int_{\beta_1}^{\beta_2} dx \; m(E) (x^2-x^{-2}+2) \\
\Sigma_2 & = & \int_{\beta_1}^{\beta_2} dx \; m(E) \frac{(1-x^2)^2}{x^2} \\
\label{Sigma}
\end{eqnarray}

where $E$ is electron energy and
\begin{eqnarray}
m(E) & = & \frac{N_e(E)}{E^2} \\
x & = & \frac{E_{min}}{E} \\
\beta_1 & = &
\left\{
\begin{array}{cc}
1      &  E_{min}>E_2  \\
\frac{E_{min}}{E_2}  &  E_{min}<E_2
\end{array}
\right.  \\
\beta_2 & = &
\left\{
\begin{array}{cc}
1      &  E_{min}>E_1  \\
\frac{E_{min}}{E_1}  &  E_{min}<E_1
\end{array}
\right.
\label{Edefinitions}
\end{eqnarray}

We have developed efficient codes to evaluate the synchrotron and SSC polarization
for arbitrary, isotropic electron spectra in the case of perfectly ordered magnetic
fields. Our code calculates the radiation powers $P_{\perp} (\omega)$ and $P_{\parallel}
(\omega)$ from all relevant radiation mechanisms separately and then evaluates the
total polarization degree as a function of photon frequency according to Eq.
(\ref{Pidef}).

\subsection{Blazar Models}

We apply the polarization code described above to both leptonic and hadronic,
stationary single-zone models of Fermi-detected blazars, as described in detail
in \cite{Boettcher13}. The blazar models have been used to produce SED fits to
a number of blazars of all sub-classes. The particle distributions and other
model parameters (in particular, the magnetic field B) required by the models 
to reproduce an individual blazar's SED are then used to evaluate the 
photon-energy-dependent degree of polarization with the code described here. 

In the leptonic blazar model, the high energy component of the SED has contributions
from synchrotron-self-Compton (SSC) and external Compton (EC) radiation, whose 
seed photons are from direct accretion disk emission, accretion disk emission reprocessed 
by the Broad Line Region (BLR) and an isotropic external radiation field. Since the 
EC high-energy emission is expected to be un-polarized, we only need to evaluate the
synchrotron and SSC polarization, and we obtain the total polarization degree in the 
leptonic model as
\begin{equation}
\Pi (\omega) = \frac{P_{\perp}^{\rm sy + SSC} (\omega) - P_{\parallel}^{\rm sy + SSC} (\omega)}
{P_{\perp}^{\rm sy + SSC} (\omega) +P_{\parallel}^{\rm sy + SSC} (\omega)} \cdot 
\frac{P_{\rm sy} (\omega) + P_{\rm SSC} (\omega)}{P_{\rm Total} (\omega)}.
\label{Pilepton}
\end{equation}

In the hadronic model, the high-energy emission consists primarily of contributions 
from proton synchrotron emission and synchrotron emission from secondary pairs produced
in cascade processes initiated by photo-pion production. At lower energies, especially 
in the X-ray range, the SSC emission of the primary electrons (primarily responsible 
for the low-frequency synchrotron peak) may also make a non-negligible contribution. 
This model only evaluates the spectra of the final decay products of the photo-pion 
production, neglecting the emission from intermediate decay products such as muons.
Thus, the polarization is calculated taking into account contributions from synchrotron 
emissions of primary electrons and protons and from secondary electron-positron pairs, 
as well as SSC from primary electrons. We point out that, even if muon and pion
synchrotron emission makes a substantial contribution to the $\gamma$-ray emission,
such a component would be expected to be equally highly polarized as the proton and
pair synchrotron emission considered in the model. Thus, a variant of the hadronic 
model that produces the same SED, but including non-negligible contributions from 
muon and pion synchrotron emission, is expected to yield very similar polarization 
signatures as calculated here. Even though the target (synchrotron) photon field 
for photopair and photopion production is polarized in the configuration considered
here, our code neglects any potential dependence of the relevant p$\gamma$ cross 
sections on the polarization of the target photon field, as such dependence is
very poorly understood and due to the assumed isotropy of the proton distribution,
its effects are expected to be negligible.

\section{\label{results}Results and Discussion}

We have applied the method of calculating the frequency-dependent degree of
polarization to a substantial number of Fermi-detected FSRQs, LBLs, IBLs, and
HBLs. We have evaluated the degree of polarization between 0.1~keV and 500~MeV, 
encompassing the energy range in which X-ray polarimeters and {\it Fermi}-LAT have 
realistic prospects of measuring high-energy polarization. We remind the reader
that the results shown here are upper limits to the polarization, assuming a
perfectly ordered magnetic field perpendicular to the line of sight (in the
co-moving frame of the emission region).

\begin{figure}[ht]
\plottwo{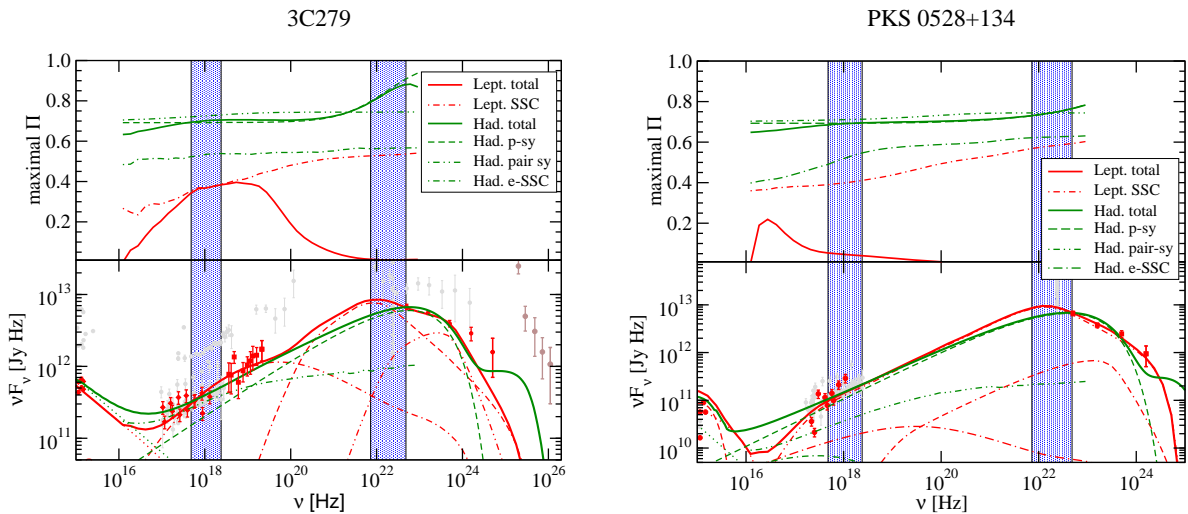}{PKS0528polarization.eps}
\caption{UV through $\gamma$-ray SEDs (lower panels) and maximum degree of
polarization (upper panels) for the two FSRQs 3C279 (left) and PKS 0528+134
(right). Leptonic model fits are plotted in red, hadronic models in green.
Different line styles indicate individual radiation components, as labeled
in the legend. Shaded areas indicate the 2 -- 10~keV X-ray range (X-ray polarimeters)
and the 30 -- 200~MeV range, in which $\gamma$-ray emission may be measurable by 
{\it Fermi}-LAT. See the electronic edition of the Journal for a color 
version of this figure.\label{FSRQpol}}
\end{figure}

\subsection{\label{LSPs}Low-Synchrotron-Peaked Blazars}

Figure \ref{FSRQpol} shows the results of SED fitting \citep[from][lower panels]{Boettcher13}
and the photon-energy-dependent degree of polarization (top panels) throughout the X-ray 
and $\gamma$-ray regime, for two representative FSRQs. In the case of FSRQs, the 
synchrotron emission from electrons (i.e., primary electrons in the hadronic model) 
does generally not contribute appreciably in the X-ray (or higher-energy) range. In 
leptonic models, the high-energy emission is typically dominated by SSC throughout 
the X-ray regime, while at $\gamma$-ray energies, EC tends to dominate.  In hadronic
models, the high-energy emission of LSP blazars is typically well reproduced by models
strongly dominated by proton synchrotron emission. 

The first, obvious result to be seen in Fig. \ref{FSRQpol} is that leptonic models
predict systematically lower degrees of polarization than (synchrotron-dominated)
hadronic models. The SSC process reduces the polarization of the synchrotron seed
photons to values typically not exceeding $\Pi_{\rm SSC} \lesssim 40$~\%, while
the proton synchrotron emission may be polarized up to $\Pi_{\rm SSC} \sim 70$ -- 75~\%
(in agreement with Eq. \ref{Pipowerlaw} for proton spectral indices $p \sim 2$ -- 3),
with the polarization gradually increasing due to the generally convex shape of the
$\gamma$-ray SED (increasing $p$ implying increasing $\Pi$). Furthermore, due to the 
transition from SSC to EC from the X-ray to the $\gamma$-ray regime in leptonic models, 
the degree of polarization is expected to decrease rapidly with photon energy, and 
vanish in the {\it Fermi}-LAT energy range. 

\begin{figure}[ht]
\plottwo{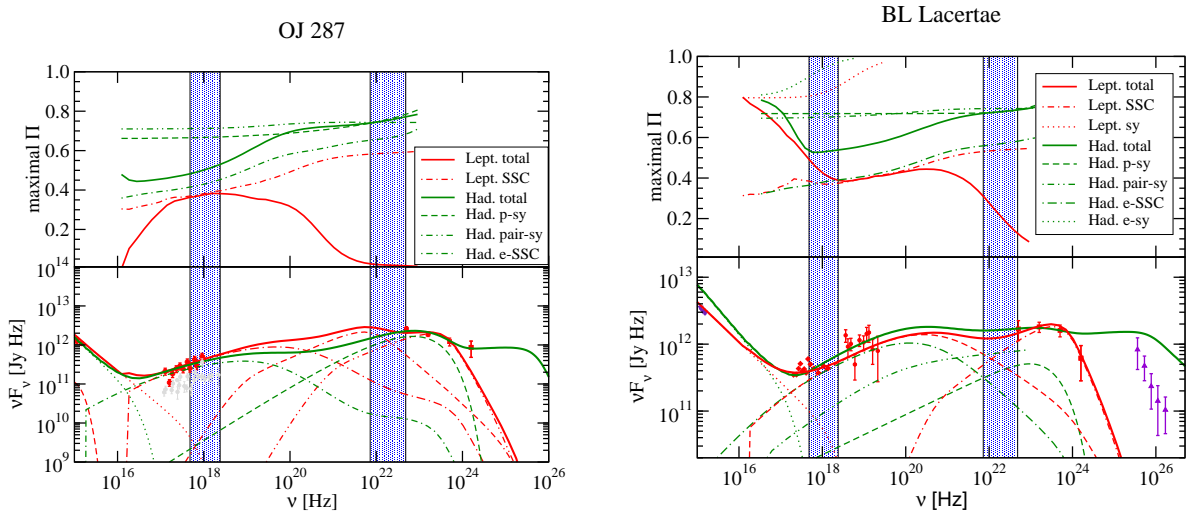}{BLLacpolarization.eps}
\caption{UV through $\gamma$-ray SEDs (lower panels) and maximum degree of
polarization (upper panels) for the two LBLs OJ~287 (left) and BL~Lacertae
(right). Same symbol/color coding as in Fig. \ref{FSRQpol}. See the 
electronic edition of the Journal for a color version of this figure.
\label{LBLpol}}
\end{figure}

Some LBLs exhibit very similar X-ray and $\gamma$-ray polarization signatures
as those discussed for FSRQs above. However, the hadronic fits of \cite{Boettcher13}
to several LBLs require their X-ray emission to be electron-SSC dominated, as in
the leptonic-model case. Figure \ref{LBLpol} shows two such examples. In that case, 
the degree of X-ray polarization predicted by hadronic models is substantially
lower than in the FSRQ case and only slightly higher than predicted by leptonic
models. In some cases (e.g., BL~Lacertae, see Fig. \ref{LBLpol}, right), the 
X-ray emission also contains a non-negligible contribution from electron-synchrotron 
radiation, which may increase the expected maximum degree of polarization, especially 
in the leptonic model, and thereby further decrease the difference between the leptonic 
and hadronic model predictions. At $\gamma$-ray energies, the same drastic difference 
between the predicted degrees of polarization persists: Hadronic models predict up to
$\sim 70$ -- 75~\% maximum $\gamma$-ray polarization, with substantially lower 
$\gamma$-ray polarization predicted by leptonic models.

\begin{figure}[ht]
\plottwo{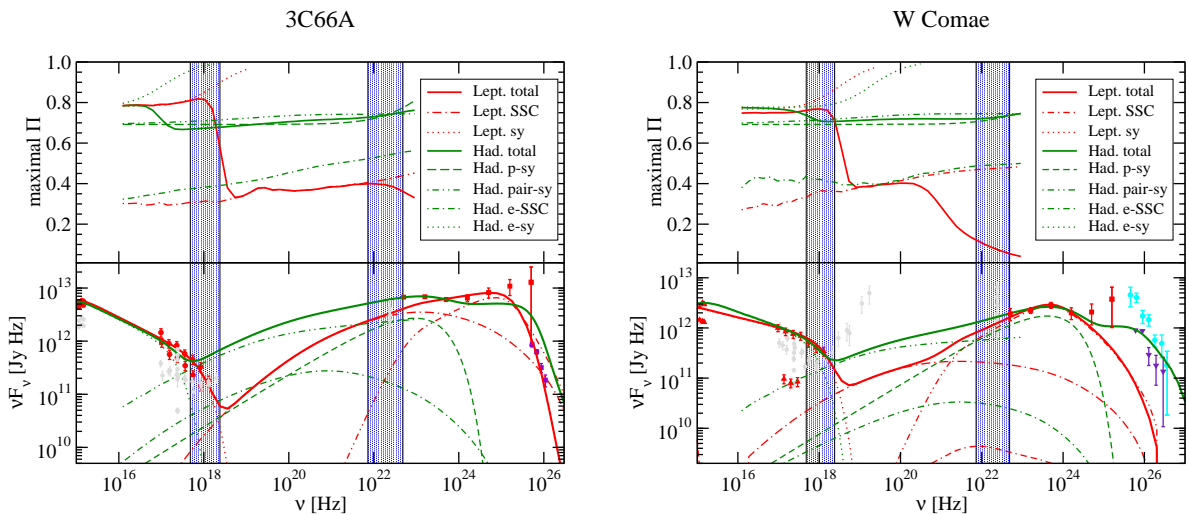}{WComaepolarization.eps}
\caption{UV through $\gamma$-ray SEDs (lower panels) and maximum degree of
polarization (upper panels) for the two IBLs 3C66A (left) and W Comae
(right). Same symbol/color coding as in Fig. \ref{FSRQpol}. See the 
electronic edition of the Journal for a color version of this figure.
\label{IBLpol}}
\end{figure}

\subsection{\label{ISPs}Intermediate-Synchrotron-Peaked Blazars}

Figure \ref{IBLpol} shows the results of SED fitting \citep[also from][]{Boettcher13}
and frequency-dependent high-energy polarization for two representative IBLs. In the
case of IBLs, the X-ray regime often covers the transition region from synchrotron
(i.e., primary electron-synchrotron in hadronic models) emission to Compton emission
in leptonic models and proton-induced emission in hadronic models. Therefore, at
soft X-ray energies, both leptonic and hadronic models can exhibit very high degrees
of polarization, dominated by the steep high-energy tail of the low-frequency synchrotron
component. Leptonic models reproduce the hard X-ray through $\gamma$-ray emission 
typically with SSC dominated emission, although in the high-energy $\gamma$-ray 
regime ($E \gtrsim 100$~MeV), an additional EC component is often required 
\citep[e.g.,][]{Acciari09,Abdo11}. Consequently, the degree of polarization is 
expected to decrease rapidly with energy to maximum values of typically $\sim 30$~\%
throughout the hard X-ray to soft $\gamma$-ray band, and may decrease even further
if the HE $\gamma$-ray emission is EC dominated. Hadronic models often require
contributions from proton-synchrotron, pair synchrotron, and primary-electron 
SSC emission. This SSC contribution slightly lowers the hard X-ray through
soft $\gamma$-ray polarization compared to purely synchrotron-dominated emission
(as in most LSPs), but still predicts substantially higher degrees of hard X-ray
and $\gamma$-ray polarization compared to leptonic models.

\begin{figure}[ht]
\plottwo{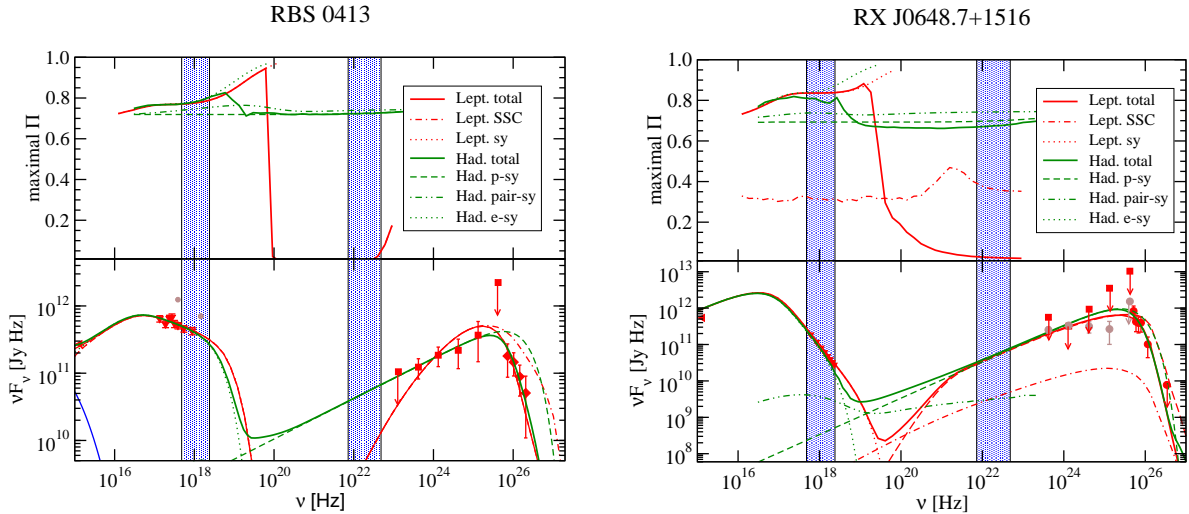}{RXJ0648polarization.eps}
\caption{UV through $\gamma$-ray SEDs (lower panels) and maximum degree of
polarization (upper panels) for the two HBLs RBS 0413 (left) and RX J0648.7+1516
(right). Same symbol/color coding as in Fig. \ref{FSRQpol}. See the electronic 
edition of the Journal for a color version of this figure.\label{HBLpol}}
\end{figure}

\subsection{\label{HSPs}High-Synchrotron-Peaked Blazars}

Two examples of SED fits and corresponding maximum polarization in the case
of HBLs are shown in Figure \ref{HBLpol}. The data and SED fits for RBS~0413
are from \cite{Aliu12}; those for RX J0648.7+1516 are from \cite{Aliu11}. In
HBLs, the X-ray emission (at least below a few 10s of keV) is typically strongly 
dominated by electron-synchrotron radiation, in both leptonic and hadronic
models. Therefore, both models make essentially identical predictions of
high maximum polarization throughout the X-ray regime. In leptonic models,
the $\gamma$-ray emission of HBLs is often well represented by pure SSC
emission, although a few cases also require additional (or even dominant)
contributions from EC \citep[see the case of RX J0648.7+1516 in the right panel
of Fig. \ref{HBLpol} for one such example;][]{Aliu11}. Combined with the very
low soft $\gamma$-ray fluxes predicted by such models, $\gamma$-ray polarization
from HBLs in leptonic models is not expected to be detectable. In hadronic 
models, the $\gamma$-ray emission of HBLs is usually well represented by
proton-synchrotron dominated scenarios, predicting a high level of maximum
polarization.

\section{\label{diagnostics}Exploiting High-Energy Polarization to Distinguish 
Leptonic from Hadronic Emission}

As pointed out, the calculations presented here assume a perfectly ordered
magnetic field and therefore only represent upper limits to the possible
degree of polarization actually observed. Therefore, in the case of a 
non-detection of X-ray and $\gamma$-ray polarization, it would seem difficult
to establish whether this is simply due to an un-ordered magnetic field,
or would actually favour a leptonic emission mechanism. However, this
ambiguity may be lifted if the degree of polarization is determined in
energy ranges of the SED that can be confidently ascribed to synchrotron
emission. 

In the case of FSRQs and LBLs, it is generally agreed that the 
near infrared (NIR) -- optical emission is dominated by synchrotron 
radiation \citep[e.g.,][]{Ghisellini98}. 
Optical polarimetry will therefore allow
one to determine the degree of order/disorder of the magnetic field in the
emission region by comparing the measured $\Pi_{\rm syn}$ with the theoretical
maximum given by Eq. (\ref{Pipowerlaw}), since the spectral index $\alpha$ of
the NIR -- optical synchrotron spectrum is easily measured. 
Due to the $\lambda^2$ dependence of Faraday rotation, the effects of
Faraday depolarization are negligible at optical wavelengths, so that the 
polarization measured here does, indeed, provide a realistic estimate of the 
degree of order of the magnetic field. Our results, as 
illustrated in Figures \ref{FSRQpol} and \ref{LBLpol} may then be re-scaled by 
that factor to arrive at a realistic prediction of the expected degree of X-ray 
and $\gamma$-ray polarization and to determine whether the expected degree of 
polarization (especially in hadronic models) is within the capabilities of 
X-ray polarimeters and {\it Fermi}-LAT. If it is, X-ray polarimetry may be
able to distinguish SSC from proton-synchrotron-dominated emission. In the 
{\it Fermi}-LAT regime, the difference between hadronic and leptonic model 
is expected to be even more obvious, and a positive detection of $\gamma$-ray
polarization by {\it Fermi}-LAT would strongly favor a hadronic emission 
scenario. 

Optical polarimetry of LSP blazars often reveals $\Pi_{\rm sy} ({\rm opt}) \sim
10$ -- 20~\%, compared to $\Pi_{\rm sy}^{\rm max} \sim 75$~\% for a typical
spectral index of $p \sim 3$. Consequently, a similar degree of X-ray and
$\gamma$-ray polarization of $\Pi_{\rm psy} \sim 10$ -- 20~\% may be expected 
for quasars in the case of hadronic emission, which is at the borderline
of the capabilities of existing polarimeters. 

In the case of IBLs and HBLs, where the X-rays (in IBLs only the soft
X-rays) are dominated by synchrotron emission, the degree of order/disorder
of the magnetic field may be determined directly from X-ray polarimetry
by compmaring the measured $\Pi_{\rm sy}$ to the theoretical maximum 
from synchrotron polarization, as evaluated by our calculations. 
As in the case of FSRQs, the resulting
re-scaling factor may be used to assess whether the expected $\gamma$-ray
polarization is within the capabilities of {\it Fermi}-LAT to determine
polarization, and if so, hadronic emission is expected to reveal itself
through a high degree of polarizaiton.

We should point out here that HBLs typically exhibit very hard photon
spectra in the {\it Fermi}-LAT range, with very low photon fluxes below
200~MeV. The detection of $\gamma$-ray polarization in HBLs therefore
seems infeasible in the near future. It has been suggested that HBLs 
may be prime targets for X-ray polarimeters since they tend to be the
brightest blazars in X-rays. However, unfortunately, leptonic and hadronic
models agree on the interpretation of the X-ray emission of HBLs as
due to electron synchrotron radiation, and therefore make identical
predictions for the X-ray polarization. Thus, X-ray polarimetry of 
HBLs is not expected to aid in distinguishing leptonic from hadronic 
emission, given current $\gamma$-ray polarimetry capabilities. Such
diagnostics would require substantially increased sensitivity and/or
$\gamma$-ray polarimetry at energies much above 200~MeV.

\section{\label{summary}Summary and Conclusions}

We have presented calculations of the maximum achievable degree of X-ray and
$\gamma$-ray polarization expected in leptonic and hadronic one-zone models
for blazars. We generally find that hadronic models predict very high degrees 
of maximum polarization for all classes of blazars, since the entire SED is 
dominated by synchrotron processes. Even when accounting for the expected 
deviations from perfectly ordered magnetic fields, the predicted degree of
polarization may be within the capabilities of current X-ray and $\gamma$-ray 
polarimeters. While in LSPs, the degree of polarization is expected 
to continually increase with increasing photon energy, it is expected to remain 
roughly constant throughout the X-ray and $\gamma$-ray regimes for ISPs and HSPs. 
Depending on the type of blazar (and, hence, the contribution that synchrotron 
emission makes to the X-ray emission), leptonic
models predict (a) moderate X-ray polarization, but vanishing $\gamma$-ray
polarization for LSPs blazars, (b) high soft X-ray polarization, rapidly
decreasing with photon energy for ISPs, and (c) high X-ray polarization
and low $< 200$~MeV $\gamma$-ray polarization (though increasing towards
higher energies beyond 200~MeV) for HSPs. 

We have outlined a method, based on optical, X-ray and $\gamma$-ray polarimetry 
that may allow us to confidently distinguish between leptonic and hadronic
X-ray and $\gamma$-ray emission in LSP and ISP blazars. Unfortunately, 
given the fact that leptonic and hadronic models make identical predictions 
for the X-ray polarization of HBLs and $\gamma$-ray polarization in those
objects is not expected to be measurable with {\it Fermi}-LAT, it is unlikely 
that X-ray polarimetry of HBLs will aid in distinguishing leptonic from hadronic 
emission. Due to the prospect of measuring the degree of order/disorder of the
magnetic field in ISPs through a comparison of X-ray polarimetry with our synchrotron
predictions, they might be the best candidates to identify hadronic processes. 
These objects will allow us to apply the method described above directly, without 
relying on optical/NIR measurements, and - unlike in HSPs - a measurement of the 
expected $\gamma$-ray polarization may be feasible using Fermi LAT or upcoming 
experiments.

An important aspect to point out is that high-energy polarization will not
be affected by Faraday rotation due to the $\lambda^2$ dependence of this 
effect. While at radio wavelengths, Faraday rotation often substantially
alters not only the orientation of the polarization direction, but may
also lead to Faraday depolarization, this effect can be ignored in X-rays
and $\gamma$-rays, thus revealing the true, intrinsic degree of order of
the magnetic field and its directionality.

Alternative ways of distinguishing leptonic from hadronic emission 
scenarios for blazars rely on the potential detectability of neutrino
emission expected in hadronic models \citep[e.g.,][]{mb92,mp01}
and characteristic variability signatures, in particular uncorrelated
synchrotron and high-energy variability, which is difficult to explain
in leptonic models, but may result more naturally in hadronic ones
\citep{Dimitrakoudis12,sw12}. However, given the sensitivity of current 
and currently planned neutrino detectors, it is unclear whether neutrino 
signals from blazars can be detected in the forseeable future. Furthermore,
also leptonic model interpretations have been suggested to explain uncorrelated 
synchrotron (optical -- X-ray) and $\gamma$-ray variability patterns
\citep[e.g.,][]{gutierrez06}, so that even such uncorrelated variability
can not be considered a unique diagnostic for hadronic emission.

Finally, we point out that our method is not limited to blazars, but may 
be applied to a large variety of X-ray and $\gamma$-ray sources as long as 
SSC emission is produced in the Thomson regime. The high energy emission
from many other X-ray and $\gamma$-ray sources (e.g., non-blazar AGN, 
gamma-ray bursts, microquasars, supernova remnants, ...), consists of 
non-thermal synchrotron and inverse Compton emission to which our
existing code can be readily applied. 
In future work, we plan to extend our calculations to multi-zone geometries, 
in which anisotropic particle distributions and arbitrary magnetic-field 
configurations can be included. Using this code will allow us to verify the scaling
arguments used to estimate the realistically expected degree of polarization
in not perfectly homogeneous magnetic fields applied above, 
and to investigate the effects of varying magnetic-field geometries
during $\gamma$-ray outbursts, as suggested by correlated optical -- $\gamma$-ray
flaring events in several blazars \citep[e.g.,][]{Marscher08}.

\section*{Acknowledgments}

We thank the referee for a helpful and constructive report.
This work was funded by NASA through Fermi guest investigator grant NNX12AP20G.
M.B. acknowledges support from the South African Department of Science and 
Technology through the National Research Foundation under NRF SARChI Chair
grant No. 64789.

\clearpage


\begin{thebibliography}{}

\bibitem[Acciari et al.(2009)]{Acciari09}
Acciari, V. A., et al., 2009, ApJ, 707, 612

\bibitem[Abdo et al.(2011)]{Abdo11}
Abdo, A. A., et al., 2011, ApJ, 726, 43

\bibitem[Aharonian(2000)]{aharonian00}
Aharonian, F. A., 2000, New Astronomy, 5, 377

\bibitem[Aliu et al.(2011)]{Aliu11}
Aliu, E., et al., 2011, ApJ, 742, 127

\bibitem[Aliu et al.(2012)]{Aliu12}
Aliu, E., et al., 2012, ApJ, 750, 94

\bibitem[Beilicke et al.(2012)]{Beilicke12}
Beilicke, M. G., et al., 2012, Proc. SPIE 8507, Hard X-Ray, Gamma-Ray and Neutron
Detector Physics XIV, 85071D

\bibitem[B\"ottcher(2010)]{Boettcher10}
B\"ottcher, M., 2010, in proc. of "Fermi Meets Jansky", Eds. T. Savolainan, E. Ros,
R. W. Porcas, \& A. Zensus; p. 41

\bibitem[B\"ottcher et al.(1997)]{bms97}
B\"ottcher, M., Mause, H., \& Schlickeiser, R., 1997, A\&A, 324, 395

\bibitem[B\"ottcher et al.(2013)]{Boettcher13}
B\"{o}ttcher, M, Reimer, A., Sweeney, K., and Prakash, A., 2013, ApJ, 768, 54

\bibitem[Bonometto et al.(1970)]{BCS70}
Bonometto, S., Cazzola, P., Saggion, A., 1970, A\&A, 7, 292

\bibitem[Bonometto \& Saggion(1973)]{BS73}
Bonometto, S., Saggion, A., 1973, A\&A, 23, 9

\bibitem[B\"uhler et al.(2010)]{Buehler10}
B\"uhler, R., et al., 2010, talk at "SCINEGHE 2010 Trieste"\footnote{\tt
http://www.rolfbuehler.net/index\_talks.html}

\bibitem[Dean et al.(2008)]{Dean08}
Dean, A. J., et al., 2008, Science, 321, 1183

\bibitem[Dimitrakoudis et al.(2012)]{Dimitrakoudis12}
Dimitrakoudis, S., Mastichiadis, A., Protheroe, R. J., \& Reimer, A., 2012,
A\&A, 546, 120

\bibitem[Forot et al.(2008)]{Forot08}
Forot, M., Laurent, P., Grenier, I. A., Gouiffs, C., \& Lebrun, F., 2008,
ApJL, 688, L29Dimitrakoudis

\bibitem[Ghisellini et al.(1998)]{Ghisellini98}
Ghisellini, G., Celotti, A., Fossati, G., Maraschi, L., \& Comastri, A., 1998, MNRAS, 301, 451

\bibitem[Gutierrez et al.(2006)]{gutierrez06}
Gutierrez, K., et al., 2006, ApJ, 644, 742

\bibitem[Krawczynski(2012)]{Krawczynski12}
Krawczynski, H., 2012, ApJ, 744, 30

\bibitem[Mannheim \& Biermann(1992)]{mb92}
Mannheim, K. \& Biermann, P. L., 1992, A\&A, 253, L21

\bibitem[Marscher et al.(2008)]{Marscher08}
Marscher, A. P., et al., 2008, Nature, 452, 966

\bibitem[M\"ucke \& Protheroe(2001)]{mp01}
M\"ucke, A., \& Protheroe, R. J., 2001, Astroparticle Physics, 15, 121

\bibitem[Orsi(2011)]{Orsi11}
Orsi, S., 2011, Astrophys. and Space Science Transactions, 7, 43

\bibitem[Pearce et al.(2012)]{Pearce12}
Pearce, M., et al., 2012, pres. at IEEE Nuclear Science Symp. 2012 (arXiv:1211.5094)

\bibitem[Poutanen(1994)]{Poutanen94}
Poutanen, J., 1994, ApJS, 92, 607

\bibitem[Rybicki \& Lightman(1985)]{RL85}
Rybicci, G. B., Lightman, A. P, 1985, \emph{Radiative processes in Astrophysics}, Wiley-VCH

\bibitem[Spanier \& Weidinger(2012)]{sw12}
Spanier, F., \& Weidinger, M., 2012, IJMP Conf. Ser., 8, 293

\bibitem[Swank et al.(2010)]{Swank10}
Swank, J., et al., 2010, "X-Ray Polarimetry: A New Window in Astrophysics", eds.
R. Bellazzini, E. Costa, G. Matt, \& G. Tagliaferri, Cambridge Univ. Press, p. 251

\bibitem[Tajima et al.(2010)]{Tajima10}
Tajima, H., et al., 2010, "X-Ray Polarimetry: A New Window in Astrophysics", eds.
R. Bellazzini, E. Costa, G. Matt, \& G. Tagliaferri, Cambridge Univ. Press, p. 275

\end{thebibliography}
\end{document}